\documentclass[%
 aip,
 amsmath,amssymb,
 reprint,%
]{revtex4-2}

\usepackage{graphicx}
\usepackage{dcolumn}
\usepackage{bm}
\usepackage[colorlinks=true,allcolors=blue]{hyperref}
\usepackage{xcolor}
\usepackage{cancel} 
\usepackage{fancyhdr}

\fancypagestyle{specialfooter}{%
  \fancyhf{}
  
  \fancyfoot[C]{Distribution A: Approved for public release; Distribution unlimited}
}

\newcommand{\update}[1]{\textcolor{black}{#1}}
\newcommand*{\rom}[1]{\expandafter\@slowromancap\romannumeral #1@}

\begin{document}


\title{High-speed plasma measurements with a plasma impedance probe}

\author{J. W. Brooks}
\email[The author to whom correspondence may be addressed: ]{jwbrooks0@gmail.com}
\author{E. M. Tejero}%
\author{M. C. Palliwoda}
\author{M. S. McDonald}%
\affiliation{ 
U.S. Naval Research Laboratory, Washington, D.C., USA
}%

\date{\today}

\begin{abstract}

	Plasma impedance probes (PIPs) are a type of RF probe that primarily measure electron density.  This work introduces two advancements: a streamlined analytical model for interpreting PIP-monopole measurements and techniques for achieving $\geq 1$ MHz time-resolved PIP measurements.  The model's improvements include introducing sheath thickness as a measurement and providing a more accurate method for measuring electron density and damping.  The model is validated by a quasi-static numerical simulation which compares the simulation with measurements, identifies sources of error, and provides probe design criteria for minimizing uncertainty.  The improved time resolution is achieved by introducing higher-frequency hardware, updated analysis algorithms, and a more rigorous approach to RF calibration.  Finally, the new model and high-speed techniques are applied to two datasets: a 4 kHz plasma density oscillation resolved at 100 kHz with densities ranging between $2 \times 10^{14}$ to $3 \times 10^{15}$~m$^{-3}$ and a 150 kHz oscillation resolved at 4~MHz with densities ranging between $4 \times 10^{14}$ to $6 \times 10^{14}$~m$^{-3}$.



\end{abstract}

\maketitle


\section{Introduction}


	Plasma impedance probes (PIPs) are a type of in-situ radio-frequency (RF) probe~\cite{rafalskyi2015, kim2016} that primarily measures plasma density by analyzing the coupled electrical impedance of the probe and the plasma.  
	The main advantages of PIPs over conventional probes, e.g. Langmuir, are that fewer assumptions and models are required to interpret their measurements; this results in smaller errors and uncertainties.  
	The main disadvantages of PIPs are that its operation requires relatively sophisticated and expensive hardware, careful calibration, and some knowledge of RF-engineering.  
	
	PIPs were originally developed~\cite{jackson1959,takayama1960,harp1964} in the 50s and 60s and have had several names over the years, including impedance, resonance, and cutoff probes.
	Since their inception, PIPs have been used in a wide range of applications:
	sounding rockets\cite{jackson1959,steigies2000,spencer2008,suzuki2010,barjatya2013,patra2013,spencer2019}, 
	DC discharges~\cite{sen1973,gillman2018}, 
	arc discharges~\cite{basu1975}, 
	plasma processing~\cite{kokura1999},
	Hall thrusters~\cite{bilen1999,hopkins2014}, 
	satellites\cite{oya1979, oya1986}, 
	and on the International Space Station \cite{wright2008,barjatya2009}.
	Physically, PIPs have been used in a number of antenna geometries including: 
	spherical monopoles~\cite{balmain1966,blackwell_measurement_2005,walker2008,hopkins2014},
	dipoles~\cite{balmain1964,balmain1969,nikitin2001,oberrath2014,rafalskyi2015,dubois2021}, 
	planar probes~\cite{sen1973}, and 
	integrated into the plasma's discharge electrodes~\cite{gillman2018}. 
	
	In their standard mode of operation~\cite{buckley1966, balmain1966,kim2004,blackwell_characteristics_2005, blackwell_measurement_2005}, PIPs electrically couple with the surrounding plasma which introduces two resonances into the probe's electrical impedance: the higher ($\omega_+$) associated with the plasma frequency ($\omega_p = 2 \pi f_p$) and the lower ($\omega_-$) with the PIP's sheath.  The PIP's impedance can be measured with a vector network analyzer (VNA) or similar instrument.  Traditionally, the plasma frequency was identified as $\omega_p \approx \omega_+$ despite known discrepancies~\cite{blackwell_measurement_2005}, and then  related to the electron density, $n_e$, using a cold plasma assumption, 
	\begin{equation} \label{eq:density}
		\begin{split}
			\omega_p =   \sqrt{\frac{n_e e^2}{m_e \epsilon_0}}.
		\end{split}
	\end{equation}
	Here, $m_e$ is the electron mass, $e$ is the fundamental charge, and $\epsilon_0$ is the vacuum permittivity.  
	The measurable density range with PIPs is traditionally considered to be roughly between $10^{10}$ to $10^{16}$ m$^{-3}$ (i.e. $1\text{ MHz} \; {\scriptstyle \lesssim} \; f_p \; {\scriptstyle \lesssim} \; 1\text{ GHz}$) where the upper limit can likely be expanded with faster hardware and a smaller probe geometry and the lower limit exceeded with larger probe geometry.   
	In addition to density, PIPs can provide the electron damping rate~\cite{spencer2008,you2016} which  includes both collisional and collisionless damping terms~\cite{walker2006, oberrath2018} (e.g. RF absorption), typically by measuring the width of the $\omega_+$ resonance~\cite{hopkins2014}.  
	However, isolating the individual components with the damping term is still an active area of research.  
	
	PIPs have recently been adapted to provide continuous, time-resolved measurements~\cite{spencer2019,dubois2021} with 10-100 kHz acquisition rates.    
	Prior to PIPs, other diagnostics have been used for continuous, time-resolved density measurements with the most common being variations on the Langmuir probe (LP): the fast-swept LP~\cite{lobbia2010,hippler2020} and the triple LP~\cite{qayyum2013, giannetti2020}.  LPs are ideal because they are `relatively' easy to setup, provide multiple plasma properties, and have time resolutions in the MHz with proper calibration.  However, their measurements typically have uncertainties between 10 to 50\% \cite{lobbia_2017}.  
	Nonintrusive optical and microwave methods also exist including interferometry~\cite{Deng2006}, spectroscopy~\cite{Wang2013,Meier2018}, and reflectometry~\cite{Sabot2006} techniques. 
	The advantages of these methods are that they do not perturb the plasma and can potentially be very accurate.  Their disadvantages are that they are not localized measurements and that they require non-trivial plasma models and assumptions to analyze.  

	

	The work presented here provides two main contributions over previous time-resolved PIP methods.  
	First, we introduce an improved PIP-monopole model (Sec.~\ref{sec:model}) that better captures electron density and electron damping rate and also introduces sheath thickness as a third measurement.  To complement this model, we leverage COMSOL simulations to both identify errors in the model and provide PIP design recommendations for minimizing these errors.  
	Second, our technique improves the time resolution of density, electron damping, and sheath thickness measurements to low MHz (Sec.~\ref{sec:methods}) through a combination  of higher-frequency hardware, rigorous calibration techniques, and improved algorithms.  
	To illustrate these advancements, the final section (Sec.~\ref{sec:results}) shows the results of the new model and techniques applied to two high-speed PIP measurements.

\section{Models \label{sec:model}}


	To extract meaningful results from a PIP, we require a model to interpret its measurements.  Below, we introduce our PIP-monopole design, our updated analytical model, a correction for the PIP's stem (the intact region of the coax cable between the probe and the final connector), and a supporting error analysis performed with COMSOL simulations.
	
	\subsection{PIP monopole antenna \label{subsec:PIP_design}}
	
\begin{figure}
			\includegraphics[]{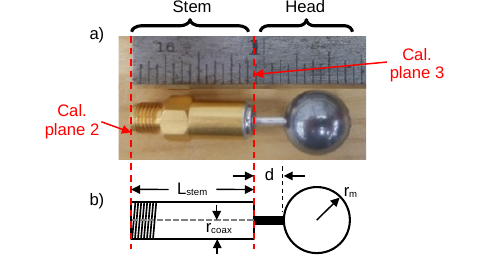}
			\caption{\label{fig:pip_model}  a) An actual PIP-monopole antenna, and b) its diagram with key dimensions indicated. }
		\end{figure}

		Figure~\ref{fig:pip_model} shows the PIP monopole antenna used in this work and its four primary dimensions: $L_{stem}$ is the length of the stem, $r_{coax}$ is the radius of the coax, $d$ is the gap between the end of the coax and the head, and $r_m$ is the radius of the monopole's head.  We fabricated the PIP from a Pasternack RG401 semi-rigid coax cable where the outer conductor and dielectric were trimmed back, and an $r_m=$ 6.35 mm (0.25 in) radius aluminum ball was drilled and press fit onto the exposed inner conductor.  The stem is defined as the length from the SMA connector to the exposed inner conductor.  The head includes the exposed inner conductor and the ball (or sphere).  
	
	\subsection{Analytical model of the PIP's head \label{subsec:PIPs_head}}
			
		Similar to previous work~\cite{buckley1966, balmain1966, blackwell_characteristics_2005, blackwell_measurement_2005, blackwell2015}, we begin our model by making a quasi-static assumption (i.e. ignoring inductive effects) and modeling the PIP's head as one or more capacitors in series.  A more complete list of assumptions is discussed at the end of this section.
		
\begin{figure}
			\includegraphics[]{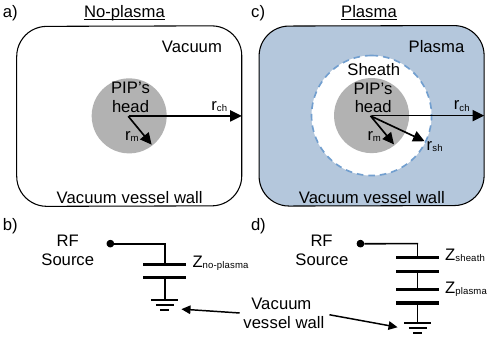}
			\caption{\label{fig:pip_model_v6b}  Simplified representations (models) of the PIP-monopole: a) simplified physics model without plasma, b) simplified electrical model without plasma, c) simplified plasma model with plasma, and d) simplified electrical model with plasma.}
		\end{figure}

		In vacuum, we model the PIP's head as an electrode of radius, $r_m$, within a grounded vacuum chamber of approximate radius, $r_{ch}$, where $r_{ch} \gg r_m$ (Figure~\ref{fig:pip_model_v6b}a). 
		When voltage is applied to the PIP, electric field lines connect the PIP's head to the wall through vacuum, and therefore we electrically model this system  as a spherical, vacuum-filled capacitor with capacitance, $C = 4 \pi \epsilon_0 / ( 1/r_m - 1/r_{ch} )$ (Figure~\ref{fig:pip_model_v6b}b). 
		The electrical impedance spectrum of any capacitor is
		\begin{equation} \label{eq:Z_cap} 
		Z(\omega) = \frac{1}{j\omega C},
		\end{equation}
		and therefore the no-plasma impedance model is		
		\begin{equation} \label{eq:Z_noplasma}
			\begin{split}
				Z_{no\mbox{-}plasma} & =\frac{1}{4 \pi j \omega \epsilon_0}\left( \frac{1}{r_m} - \frac{1}{r_{ch}} \right) \\
				&  \approx  \frac{ Z'}{j \omega'}. 
			\end{split}
		\end{equation}
		Here, $j$ is the imaginary unit, and we have defined a normalized frequency, $\omega' \equiv \omega / \omega_p$, and a characteristic impedance, $Z' \equiv 1/(4 \pi \epsilon_0 r_m \omega_p)$.    
		
		When plasma is present, a sheath of radius, $r_{sh}$, and thickness, $t_{sh}=r_{sh}-r_m$, forms between the sphere and plasma where $r_m < r_{sh}\ll r_{ch}$ as shown in Figure~\ref{fig:pip_model_v6b}c.  
		We electrically model the sheath and plasma regions as two spherical capacitors in series (i.e. concentric) as shown in Figure~\ref{fig:pip_model_v6b}d.   
		
		For the sheath region, we assume that its density is much lower than the plasma region, and therefore we model the sheath as a homogeneous vacuum ($\epsilon \approx \epsilon_0$).  Using the spherical capacitor model, the sheath's impedance is 
		\begin{equation} \label{eq:Z_sheath}
			\begin{split}
				Z_{sheath}  &= \frac{1}{4 \pi j \omega \epsilon_0}\left( \frac{1}{r_m} - \frac{1}{r_{sh}} \right) \\
				&  =  \frac{Z'}{j\omega'} t_{sh}'
			\end{split}
		\end{equation}
		where we have defined a normalized sheath thickness, $t_{sh}' \equiv t_{sh}/r_{sh}$, 
		where $t_{sh}'$ is bounded between 0 and 1.
		
		For the plasma region, we assume it to be cold, collisional, homogeneous, and non-magnetized, and therefore its relative permittivity~\cite{blackwell_measurement_2005} is
		\begin{equation} \label{eq:epilson_r}
			\begin{split}
				\epsilon_{p} &=  \left(1 - \frac{1 }{\omega' \left( \omega' - j \nu' \right) } \right), \\
			\end{split}
		\end{equation}
		where $\nu' \equiv \nu / \omega_p$ and $\nu$ is the electron damping rate.
		Modeling the plasma region as the outer spherical capacitor, its impedance is
		\begin{equation} \label{eq:Z_plasma}
			\begin{split}
				Z_{plasma} &  = \frac{1}{4 \pi j \omega  \epsilon_0 \epsilon_{p}}\left( \frac{1}{r_{sh}} - \frac{1}{r_{ch}} \right) \\
				& \approx  \frac{Z'}{j \omega'} \frac{1}{\epsilon_p} \left(1 - t_{sh}' \right). \\ 
			\end{split}
		\end{equation}

		
		
		
		

		
		Combining the sheath and plasma impedance, we solve for the head's total impedance in the presence of plasma,
		
		
		\begin{equation} \label{eq:Z_total2}
			\begin{split}
				Z_{tot}  &=  Z_{sheath} + Z_{plasma}\\
				&=  \frac{Z'}{j \omega'} \left( t_{sh}' +   \frac{\left(1 - t_{sh}'\right)}{\epsilon_p} \right)\\
			\end{split}
		\end{equation}
		
		
	
		
		\noindent which has three unknown parameters: plasma frequency ($\omega_p$), electron damping rate ($\nu$), and sheath thickness~($t_{sh}$) which can be determined by fitting Equation~\ref{eq:Z_total2} to calibrated measurements.
		
		A\update{n alternate and} simpler method of determining these parameters is to identify resonances in the calibrated measurements and relate them to the predicted resonances in the model. 
		Resonances, in any electrical system, occur at the frequencies where $\text{Im}(Z)=0$.  Applying this to Eq.~\ref{eq:Z_total2}, we find that $\text{Im}(Z_{tot})$
		has two zeros,
		
		\begin{equation} \label{eq:Z_total_zeros}
			\begin{split}
				 \left| \frac{\omega_{\pm}}{\omega_p}\right|
 				 &= \sqrt{ \frac{1 + t_{sh}' -\nu'^2 \pm \sqrt{ \left(1 + t_{sh}' - \nu'^2\right)^2 - 4 t_{sh}'} }{2} }, \\
			\end{split}
		\end{equation}
		
		\noindent which are both are functions of $\omega_p$, $\nu'$ and $t_{sh}'$.  We refer to $\omega_+$ as the plasma resonance and $\omega_-$ as the sheath resonance because of their strong dependence on $\omega_p$ and $t_{sh}'$, respectively. 
		Previously, it was assumed that $\omega_+ \approx \omega_p$, but Eq.~\ref{eq:Z_total_zeros} shows that this is true only when $\nu'$ is sufficiently small\cite{blackwell2015}. 
		\update{These resonances are further discussed in a previous publication\cite{blackwell_characteristics_2005}.}
		
		Further solving, we remove the sheath's contribution from Im($Z_{tot}$) and therefore better isolate $\omega_p$ by subtracting $Z_{no\text{-}plasma}$ from $Z_{tot}$ to get, 
	
		\begin{equation} \label{eq:Z_diff2}
			\begin{split}
				Z&_{diff}  = Z_{tot} - Z_{noplasma} \\
				&=  \frac{1}{4 \pi j \omega \epsilon_0}  \left[ \frac{ \left(\frac{1}{r_{sh}} \, \text{-} \, \frac{1}{r_{ch}}\right)}{\epsilon_p}+ \left(\cancel{\frac{1}{r_m}} \, \text{-} \, \frac{1}{r_{sh}}\right) - \left(\cancel{\frac{1}{r_m}} \, \text{-} \, \frac{1}{r_{ch}}\right)\right] \\
				&\approx \frac{Z'}{j \omega'}   \left( \frac{1}{\epsilon_p }- 1\right) \left(1 - t_{sh}' \right).
			\end{split}
		\end{equation}
		
		\noindent Expanding this equation provides, 
				\begin{equation} \label{eq:Z_diff_exp}
			\begin{split}
				\frac{Z_{diff}}{ Z' (t_{sh}' - 1)}  &=   \frac{  \nu'\omega' +j\left( \omega'^2 - 1 \right)}{\omega' \left( \nu'^2 \omega'^2 + \left( \omega'^2 - 1\right)^2\right)},\\
			\end{split}
		\end{equation}
		\noindent where $\text{Im}(Z_{diff})$ has a zero at $\omega' = 1$ (i.e. $\omega=\omega_p$).  This convenient cancellation occurs because we assume that the sheath is a homogeneous vacuum; we expect a similar but diminished effect in a more realistic sheath environment. 
		
		To summarize, this simpler method consists of two steps: i) identifying $\omega_p$ as the only zero in calibrated measurements of Im($Z_{diff}$), and ii) locating the two zeros in calibrated measurements of Im($Z_{tot}$) and relating these to Eq.~\ref{eq:Z_total_zeros} to determinue $\nu$ and $t_{sh}$.  
		

		
		
		Revisiting Eq.~\ref{eq:Z_total_zeros}, we find that the two zeros (resonances) merge and then vanish when
		
		
		\begin{equation} \label{eq:collisional_limit}
			\begin{split}
				 \nu' & \geq 1 - \sqrt{t_{sh}'} . \\
			\end{split}
		\end{equation}
		The disappearance of these zeros with high damping has been previously observed~\cite{gillman2018} but not explained.  
		
\begin{figure}
			\includegraphics[]{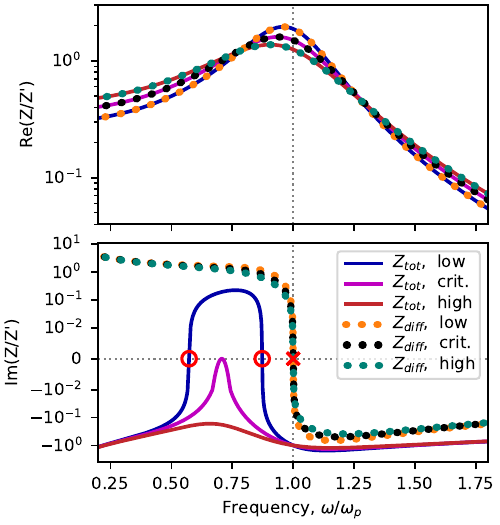}
			\caption{\label{fig:PIP_analyical_model}$Z_{tot}$ and $Z_{diff}$ are shown for its three values of $\nu' + \sqrt{T_s}$: \emph{low}, \emph{critical}, and \emph{high}.  The zeros (`o') of $\text{Im}(Z_{tot})$ merge and vanish as $\nu' + \sqrt{t_{sh}'}$ increases, but the same zero (`x') for $\text{Im}(Z_{diff})$ exists at $\omega=\omega_p$ for all three cases. }
		\end{figure}

		
		To better understand $Z_{tot}$ and $Z_{diff}$, we plot both in Figure~\ref{fig:PIP_analyical_model} for three cases of Eq.~\ref{eq:collisional_limit} : 
		\emph{low}, \emph{critical}, and \emph{high} where $\nu' + \sqrt{t_{sh}'}$ is equal to 0.9, 1.0, and 1.1, respectively, with $t_{sh}'=0.25$. 		
		Figure~\ref{fig:PIP_analyical_model}a shows that the real components of $Z_{tot}$ and $Z_{diff}$ are identical and have peaks near $\omega_p$; a relationship between the width of these peaks and $\nu$ has been previously derived~\cite{hopkins2014}.
		Figure~\ref{fig:PIP_analyical_model}b shows that the two zeros of $\text{Im}(Z_{tot})$ exist at the 	\emph{low} case, converges to a single zero at the 	\emph{critical} case, and vanish at the 	\emph{high} case.  However, the single zero at $\omega=\omega_p$ exists for all three cases in $\text{Im}(Z_{diff})$.
		
		The models presented above provide several improvements over the previous PIP-monopole models.  Specifically, 
		i) the models now include sheath thickness as a measurable quantity,
		ii) all three plasma parameters ($\omega_p$, $\nu$, and $t_{sh}$) can be determined by fitting Eq.~\ref{eq:Z_total2} to calibrated measurements or by locating the resonances in the measurements, and
		iii) the dependencies and disappearance of the zeros in PIP measurements can now be explained.  
		

		The models above include a number of assumptions which both limit their applicability and introduce error. 
		First, these models are designed with spherical geometries.  However, if we reconstruct the analytical model using planar or cylindrical geometries, $\text{Im}(Z_{diff})$ always results in a zero at $\omega=\omega_p$, suggesting that some aspects of this model are independent of PIP geometry. 
	    Second, the models assume that the plasma is cold, non-magnetized and homogeneous which might not be reasonable for all plasma environments.  
	    \update{Third, we used a linearization assumption when deriving Eq.\ref{eq:epilson_r}, which is only valid when the exponent in the Boltzmann relation, $V_{RF} / T_{eV}$, is small.  Here, $V_{RF}$ is the driven RF voltage, and $T_{eV}$ is the plasma temperature with units in eV. }
		Fourth, we make a quasi-static assumption which allows us to ignore impedance from inductance and justify it because $c/f_p$ is much larger than our dimensions.  
		Fifth, we assume that the sheath is a  homogeneous vacuum and that the transition to the plasma region is discontinuous.  In Sec.~\ref{subsec:COMSOL model}, we show that this is reasonable for thin sheaths.
		Sixth, we assume any capacitance between the PIP's head and the stem's outer conductor is small or at least comparable with the capacitance between the head and wall.  In Sec.~\ref{subsec:COMSOL model}, we show that this is reasonable for certain probe dimensions. 
		Finally, we assume that we can accurately calibrate up to the PIP's head, which includes accurately calibrating the PIP's stem as discussed in the next section.
		

	\subsection{Calibrating the PIP's stem \label{subsec:PIPs_stem}}
	
\begin{figure}
			\includegraphics[]{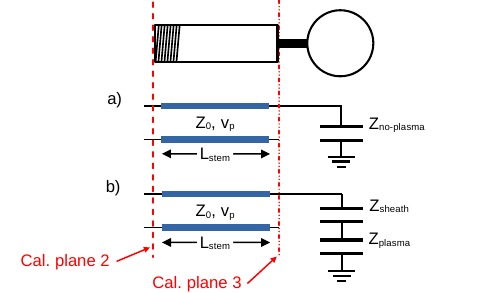}
			\caption{\label{fig:pip_stem_model} The PIP's stem is modeled as a lossless transmission line for both the a) no-plasma case and b) with-plasma case.  The PIP diagram is provided for reference. }
		\end{figure}

		Any measurement of the PIP will include impedance contributions of both the PIP's head and stem.  If ignored, the stem can introduce significant errors (e.g. 50\%) to measurements.  In this work, we remove the stem's impedance (i.e. calibrate the stem) by modeling it as a lossless transmission-line as shown in Figure~\ref{fig:pip_stem_model}.
		
		The lossless tranmission-line model~\cite{pozar2011} is
		
		
		
		
		\begin{equation} \label{eq:stem_model_Z2}
		\begin{split}
				Z_{2} = Z_0 \frac{Z_3 + j Z_0 \, \text{tan}(\omega L / v_p)}{Z_0 + j Z_3 \text{tan}(\omega L / v_p)} 
		\end{split}
		\end{equation}
		where $Z_{3}$ is the impedance of the PIP's head (at calibration plane~3), $Z_{2}$ is the combined impedance of the head and stem (at calibration plane~2), and $L=L_{stem}$.  The constants are the published values of the coaxial cable: its characteristic impedance ($Z_0$), 
		and velocity of propagation ($v_p$).   		
		We justify using this model because the stem is the unmodified portion of the original coax cable and because the stem's length is short compared with the cable's attenuation scale length.
		
		In practice, we typically solve Eq.~\ref{eq:stem_model_Z2} for $Z_3$,
		\begin{equation} 		 \label{eq:stem_model_Z3}
				\begin{split}
					Z_{3} = Z_0 \frac{j Z_0 \, \text{tan}(\omega L / v_p) - Z_2}{ j Z_2 \text{tan}(\omega L / v_p) - Z_0},
				\end{split}
		\end{equation}
		and apply this to calibrated measurements at plane~2~($Z_2$) to remove the stem.
				
\begin{figure}
			\includegraphics[]{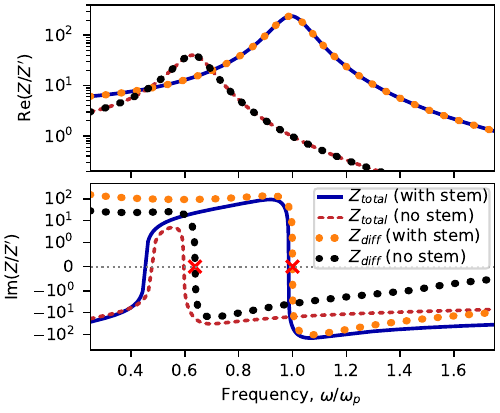}
			\caption{\label{fig:PIP_analyical_model_with_stem}  An example of simulated $Z_{tot}$ and $Z_{diff}$ is shown where the stem is and is not included.  The red ``x''s show the respective plasma frequencies.  Failing to calibrate the stem causes the true plasma frequency to be underestimated by~36\%. }
		\end{figure}

		To illustrate the importance of calibrating the stem, we simulate a plasma using Eq.~\ref{eq:Z_total2} with the following values: $\nu'=0.15$, $t_{sh}'=0.2$, and $Z' = 2250$ $\Omega$.  
		We then add the stem's impedance by applying  Eq.~\ref{eq:stem_model_Z2} to the simulated measurement with $Z_0 = 50 $  $\Omega$, $L_{stem}=21.0$ mm, and $v_p = 0.695 \, c$.
		We then performed a similar process for Eq.~\ref{eq:Z_diff2}.   Figure~\ref{fig:PIP_analyical_model_with_stem} shows the results where the stem is both included and not included. 
		In this example, the stem shifts the measured plasma frequency from $\omega  = \omega_p$  to $\omega  = 0.64 \, \omega_p$.  
		By not calibrating the stem in this example, we would introduce a 36\% error in plasma frequency and 60\% error in density to our measurements.  
		The COMSOL error analysis section below shows similar results.
		
	\subsection{COMSOL error analysis \label{subsec:COMSOL model}}
	
		In this section, we describe our COMSOL numerical model of the PIP-monopole and its application to i) validate the previous analytical models, ii) investigate sources of error in our models, and iii) provide recommendations for a PIP-monopole design that minimizes these errors.  
			
	
		Effectively, our numerical model makes the same or similar assumptions as the analytical model with the primary difference being that the numerical model is solved in three spatial dimensions with realistic boundary conditions.  
		We constructed this model in COMSOL Multiphysics $^ \text{\textregistered}$ \cite{COMSOL}, a commercial finite element solver, and solved \update{it} as described by the following steps.  
		First, we modeled the PIP antenna and stem using the geometries shown in Figure~\ref{fig:pip_model}b, where $L_{stem}=$ 0.827~in (21.0~mm), $d=$ 0.219~in (5.56~mm), $r_m=$ 0.5~in (12.7~mm), and $r_{coax}=$ 0.0725~in (1.84~mm) unless otherwise specified.  The stem's dielectric constant was $2.07$. The antenna was placed in the center of our domain: an electrically grounded 0.5~m diameter cylinder that was 1~m long.  
		Second, we filled the domain with an appropriate dielectric: $\epsilon=\epsilon_0$ for a vacuum environment and Eq.~\ref{eq:epilson_r} for \update{the} plasma environment.   We modeled the PIP's sheath as a thin, homogeneous vacuum of thickness, $t_{sh}$, around the probe's head, exposed inner coax, and the length of the stem.  Same as the analytical function, this model is a function of three parameters: $\omega_p$, $\nu'$, and $t_{sh}'$.  
		Third, \update{we used} COMSOL to solve Poisson’s equation,
		\begin{equation} \label{eq:poisson}
			\nabla \cdot \epsilon \nabla \phi = -\rho_f,
		\end{equation}
		to find the electrostatic potential, $\phi$, throughout the domain.  In this equation, $\rho_f$ is the free charge density and $\epsilon$ is the dielectric constant.  
		Fourth, we modeled the PIP's electrical impedance as a capacitor (Eq.~\ref{eq:Z_cap}) with capacitance,
		\begin{equation} \label{eq:capacitance}
			C = \frac{1}{\Delta V} \iint - \nabla \phi \cdot \hat{n} \, da.
		\end{equation}
		In this expression, $\Delta V$ is the potential difference between the probe and the chamber wall, and $\hat{n}$ is the unit vector normal to the surface of the probe.
		In the case where this model is fit to measurements, a fifth step is to use an external Python code to iteratively adjust $\omega_p$, $\nu'$, and $t_{sh}'$ and then solve steps 2 to 4 using a least-squares algorithm until the residual between the measured and simulated impedances \update{is} minimized.

\begin{figure}
			\includegraphics[]{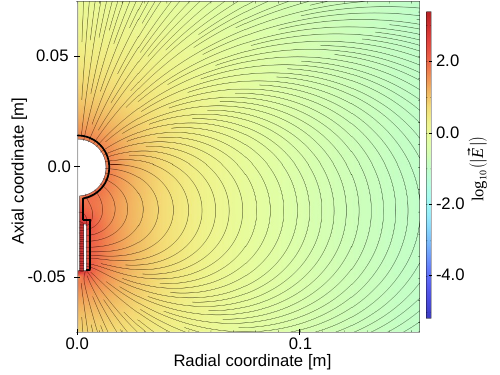}
			\caption{\label{fig:COMSOL_simulation} A COMSOL simulation of the PIP's head and stem in vacuum shows the surrounding electric field (streamlines and magnitude) solved at 100 MHz.  The electric field has units of V/m. }
		\end{figure}

		Figure~\ref{fig:COMSOL_simulation} shows a portion of the domain solved at 100 MHz and at vacuum.  The streamlines show the electric field orientation and the background color shows the field's magnitude.  The white regions are electrical conductors, and the sheath boundary is indicated by the bold, black line.  Because the field's magnitude scales roughly as $E\propto r_m^2/r^2$, the effective volume of plasma that the PIP measures is within a few radii of the PIP's sphere.   
		
\begin{figure}
			\includegraphics[width=3.33in]{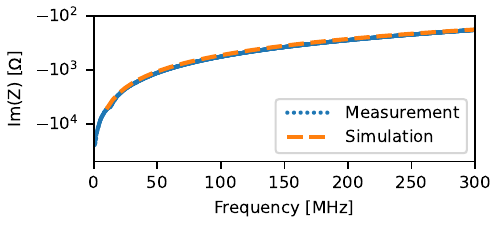}
			\caption{\label{fig:COMSOL_vac_results} COMSOL simulation results of the PIP monopole in vacuum overlayed with an actual calibrated $Z_{noplasma}$ measurement, recorded with a VNA, of the same geometries.  
			}
		\end{figure}

		Next, we compare our COMSOL model (dashed orange) to vacuum measurements (blue dots) of the PIP in 
		Figure~\ref{fig:COMSOL_vac_results}.  The COMSOL model assumes the same geometries as the physical PIP-monopole, and the two traces show good qualitative agreement.
				
\begin{figure}
			\includegraphics[width=3.33in]{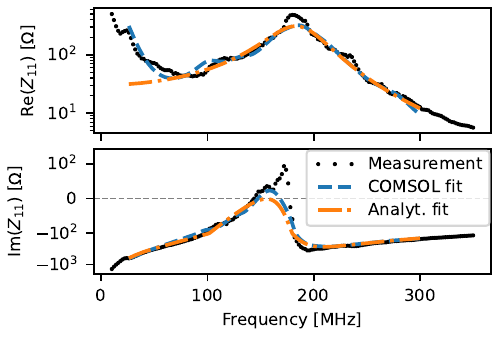}
			\caption{\label{fig:COMSOL_plasma_results} Comparison of calibrated PIP plasma measurements with least-squares fits of the COMSOL simulation and analytical model. 
			} 
		\end{figure}

		In Figure~\ref{fig:COMSOL_plasma_results}, we fit both our COMSOL and analytical models to a calibrated plasma measurement.   Both models qualitatively fit well to the gross features of the measured data, and the COMSOL fit additionally achieves some of the finer features in the measurements.  However, both models rely on simplistic assumptions (i.e. \update{there is} missing physics) and cannot fully reproduce the measurements.  It is clear that a vacuum sheath with a hard boundary is an unphysical model for the sheath, although it does allow for an accurate prediction of the sheath resonance.  In addition, the fluid treatment of the plasma in the calculation of the plasma dielectric also neglects kinetic effects in the sheath and cannot account for resonant absorption in the sheath that would be lumped into the damping rate\cite{walker2006, oberrath2018}.
		
		Next, we simulate our COMSOL model over a range of parameters:  plasma properties ($\omega_p$, $\nu'$, $t_{sh}'$), probe dimensions ($r_m$, $d$), and with and without the stem calibration technique discussed in Sec.~\ref{subsec:PIPs_stem}.  For each configuration, we generate a measurement of $Z_{diff}$, identify the plasma frequency as Im($Z_{diff}$)=0, calculate the difference between this and the plasma frequency used in the simulation and take the absolute value to get the error.  
		
\begin{figure}
			\includegraphics[width=3.33in]{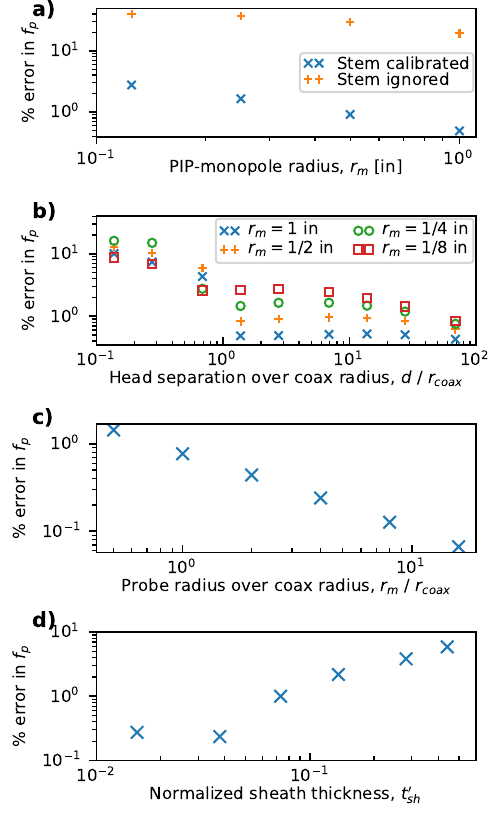}
			\caption{\label{fig:COMSOL_ant_optimize_all} Comparison percent error in determining plasma frequency between the COMSOL simulation and analytical model.  The comparison was conducted over several variables : a) calibration of the stem's transmission line (TL) over a range of probe radius, b) a range of $d/r_{coax}$ for several probe radii, c) a range of sheath thickness, and d) a range of $r_m/r_{coax}$. }
		\end{figure}

		With this method, we identified the largest source of error, between 20\% and 40\%, as failing to calibrate the stem (Figure~\ref{fig:COMSOL_ant_optimize_all}a).  This is similar to the results in Sec.~\ref{subsec:PIPs_stem}.
		
		We identified the second largest source of error, on the order of 0.5 to 15\%, to the proximity of nearby, grounded electrical conductors, most notably the PIP's stem.  When $d$, the gap distance between the PIP's stem and head is small (as shown in Figure~\ref{fig:COMSOL_ant_optimize_all}b), the capacitance between the head and the stem becomes significant which invalidates the assumptions used in Section~\ref{subsec:PIPs_head}.  A similar but smaller effect occurs when the radius of the stem increases as shown in Figure~\ref{fig:COMSOL_ant_optimize_all}c.

		

		We attribute the third largest source of error, on the order of 0.2 to 5\%, to the over-simplicity of our sheath model.  If a significant portion of the measurable plasma volume around the PIP is occupied by the sheath (i.e. the sheath is ``thick''), then the errors in the sheath model become significant.  Figure~\ref{fig:COMSOL_ant_optimize_all}d shows how this error increases with larger sheath thickness.  Because $t_{sh}$ is related to the Debye length, $t_{sh}$ is therefore also a function of density, and $r_m$ should always be chosen so that the sheath is ``thin''.  
		
		Other sources of error were identified, e.g. large damping, but are not discussed here because their errors were under 0.5 \%.  A notable omission from this section are errors from density gradients, and we plan to study these in the future.


		To minimize these errors, we have developed the following ``rules of thumb'' for designing a PIP-monopole.  First, the stem must be designed in a way that it can be accurately calibrated.  This includes building the PIP stem from materials with published dielectric properties (e.g. RG401 semirigid cable).  This also includes having a convenient connector (e.g. SMA) at the second calibration plane.  Second, all PIP geometries must be much less than $ c/f_p$ to follow the quasi-static assumption.  Third, $L_{stem}$ must be much smaller than its attenuation scale length (see  Sec.~\ref{subsec:PIPs_stem}).  Fourth, the gap, $d$, between the PIP's stem and head should be $d  \; {\scriptstyle \gtrapprox} \; r_{coax}$ and $d \; {\scriptstyle \gtrapprox} \; r_{m}$ but not too large as to cause structural issues.  Fifth, the PIP's radius, $r_m$, should be $r_m \gg t_{sh}$, $r_m \gg r_{coax}$, but also smaller than any spatial resolution requirement (e.g. density gradients).  

\section{Methods \label{sec:methods}}

	This section describes the setup, calibration, and the analysis we used to provide our high-speed PIP measurements.  
	
	\subsection{Experimental setup \label{subsec:setup}}

		The plasma source used in this work was a thermionic LaB$_6$ hollow-cathode with typical argon flowrates of order 10 sccm.  Power supplies sustained a DC plasma discharge between the cathode and  an annular anode, located roughly 1 inch downstream.  
		The vacuum chamber was a stainless steel cylinder, approximately 29 inches in diameter, 36 inches long, and electrically grounded.  During plasma operation, pumps maintained a background pressure between $10^{-5}$ to $10^{-4}$ Torr.  
		An electromagnet, installed around the cathode, created a coaxial field that fostered oscillations in the plasma.  The field strength at the PIP, approximately 10 inches downstream, was roughly 1 G.  At this field strength, the cyclotron frequency is much less than the $\omega_p$ when $\omega_p$ is of order 100 MHz.  Therefore we consider the plasma to be unmagnetized.  
		
	 	In addition to the PIP, we included two time-resolved diagnostics to provide additional measurements of the plasma oscillations: the hollow-cathode's discharge current and current from an ion saturation probe (adjacent to the PIP).

\begin{figure}
			\includegraphics[]{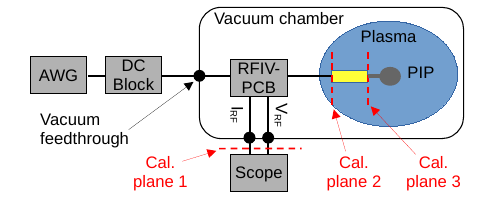}
			\caption{\label{fig:time_series_setup} Block diagram of the time-resolved PIP setup. }
		\end{figure}

		The time-resolved PIP setup is diagrammed in Figure~\ref{fig:time_series_setup}.  Here, a Tektronix AWG7062B arbitrary waveform generator (AWG) drove sequential pulses through a series of cables, feedthroughs, and circuits to the PIP where the pulses were distorted by the plasma and reflected back.  The local voltage $V_{RF}$ and current $I_{RF}$ of the outbound and reflected pulses were measured near the antenna using a custom (RF-IV) PCB (see App. A) and a Keysight DSOS054A oscilloscope at 10 GS/s/channel.  The impedance of the system was calculated using
		
		\begin{equation} \label{eq:Z_RFIV}
			\begin{split}
				Z_1(\omega) = \frac{\text{FFT}\left\{V_{RF}(t)\right\}}{\text{FFT}\left\{I_{RF}(t)\right\}}
			\end{split}
		\end{equation}
		where FFT is the fast-Fourier transform, and the 1 subscript indicates the measurement was made at calibration plane~1.
		A DC blocking filter (Mini Circuits BLK-89 S+) was included to protect the AWG from DC currents.  The majority of the RF cabling was a flexible RG316 with SMA connectors.  The outer conductor of the PIP's RF cable was grounded to the chamber at the RF feedthrough.  Care was made to route the RF cabling so that it was secure and to prevent it from being heated by the plasma and therefore cause calibration drift.   
		
		
\begin{figure}
			\includegraphics[]{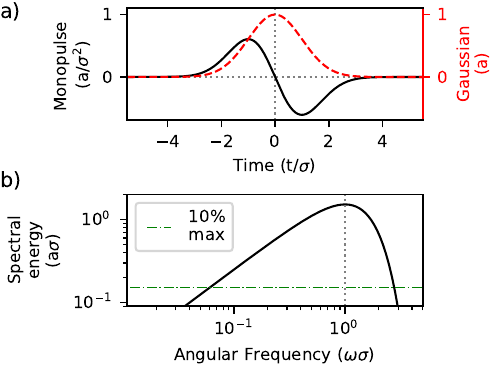}
			\caption{\label{fig:monopulse}a) Normalized Gaussian and Gaussian-monopulse waveforms.  b) Power spectrum of the monopulse with its peak at $\omega\sigma=1$ and its width is indicated by  10$\%$-of-max threshold.  }
		\end{figure}

		The pulse transmitted by the AWG was the first derivative of the Gaussian distribution, often called the Gaussian-monopulse~\cite{spencer2019,dubois2021},
		\begin{equation} \label{eq:monopuse}
			\begin{split}
				%
				g_{mp}(t; \sigma) = \frac{a t}{\sigma^2} \text{exp}\left(-\frac{1}{2}\left(\frac{t}{\sigma}\right)^2\right),
			\end{split}
		\end{equation}
		\noindent where $a$ and $\sigma$ are the amplitude and standard deviation, respectively, of the Gaussian.  
		We use this pulse because its spectral energy distribution,
		
		\begin{equation} \label{eq:monopuse}
			\begin{split}
				\left|G_{mp} (\omega; \sigma)\right| = a \sigma \omega \, \text{exp} \left( \text{-}\frac{1}{2} \sigma^2 \omega^2 \right),
			\end{split}
		\end{equation}
		can be targeted at a particular range of plasma frequencies by adjusting $\sigma$.   Specifically, the distribution's peak is at $\omega=1/\sigma$, and its width is roughly between $0.06 < \omega \sigma <2.8$ as defined by a 10\%-of-max threshold.  The monopulse, its spectral energy distribution, and the Gaussian are shown in Figure~\ref{fig:monopulse}. 
		
		The other important pulse parameter is $\tau$, the time between sequential monopulses, and this sets the time resolution of the PIP measurements. The trade-off for increasing the time resolution (i.e. decreasing $\tau$) is a lower frequency resolution (and therefore a lower density resolution) as the scope measures fewer points associated with each pulse.   The upper limit of $\tau$ is set by two factors: i) the time-decay of the reflected pulse, which is proportional to one over the damping term~\cite{dubois2021} $1/\nu$ and ii) preventing sequential reflected and outbound pulses from overlapping at the RF-IV board.     

	\subsection{RF calibration \label{subsec:calibrations}}
		
		In setting up for this experiment, we used the following calibration procedure before pumping down and taking measurements.  This discussion references the calibration planes shown in Figures~\ref{fig:pip_model},~\ref{fig:pip_stem_model}, and~\ref{fig:time_series_setup}, and the subscripts 1, 2, and 3 represent the respective measurements at the three calibration planes.  
		
		Calibration is essential because the measured impedance at the scope, $Z_1$, includes the desired impedance of the PIP's head, $Z_3$, as well as undesired contributions from everything between.  To isolate the impedance of the PIP's head, we perform a 1-port calibration method (described below) to extend the calibration plane from 1 to 2 and then use Eq.~\ref{eq:stem_model_Z3} to extend the calibration plane from 2 to 3.    
		
		
		
		\subsubsection*{Calibration with a 1-port error model}
		
		To remove the impedances between the scope and stem (cal. planes~1 to~2), we employ a 1-port error model~\cite{janjusevic_2019, dubois2021}
		\begin{equation} \label{eq:ABCD_subtract}
			\begin{split}
				Z_2 & = \frac{Z_1 - B}{ A  - C Z_1},
			\end{split}
		\end{equation}
		where $Z_1$ and $Z_2$ are the measured impedances at their respective cal. planes.   The three parameters ($A$, $B$, and $C$) are unique to a given hardware setup and include nuances such as cable temperature, length and routing.  
		
		Before applying this expression, we needed to first solve for $A$, $B$, and $C$.  We did this by using a VNA and a custom PCB standards board (App.~\ref{app:pcbs}) that contains three standards: an R, an L, and a C circuit element.  The PCB contains a fourth circuit, a resonant RLC circuit, that we used to check the efficacy of the 1-port calibration.
		
		Step 1 was to acquire ``truth'' measurements of our standards.  We do this by performing a single-port calibration of the VNA's port~1, connecting our three standards and the RLC circuit to the VNA, and measuring each impedance: $Z_{R}$, $Z_{L}$, $Z_{C}$, and $Z_{RLC}$.   
		
		Step 2, we assembled the experimental setup diagrammed in Figure~\ref{fig:time_series_setup}, including all of the cabling/connectors/feedthroughs/etc that would be present in our plasma measurements.  We then chose appropriate values for $\tau$ and $\sigma$ based on expected plasma conditions and started the AWG pulsing.  We then placed our three standards, RLC circuit, and PIP at plane~2 and measured each with the oscilloscope.  To provide a single, multi-pulse-averaged measurement for each standard, we set the scope to trigger on the outbound pulse and to average over 1000 pulses.  We then used Eq.~\ref{eq:Z_RFIV} to convert the measurements to impedance: $Z_{1,R}$, $Z_{1,L}$, $Z_{1,C}$, $Z_{1,RLC}$, and $Z_{1,noplasma}$. 
		
		Step 3, we resolved Eq.~\ref{eq:ABCD_subtract} using $Z_{1,R}$, $Z_{1,L}$, $Z_{1,C}$ and our three truth standards (substituted into $Z_2$) to get the expression,
		
		\begin{equation} \label{eq:ABCD_inverse}
			\begin{split}
				\begin{bmatrix}
					Z_{2,R} & 1 & \text{-} Z_{1,R} \, Z_{2,R} \\ 
					Z_{2,L} & 1 & \text{-} Z_{1,L} \, Z_{2,L} \\
					Z_{2,C} & 1 & \text{-} Z_{1,C} \, Z_{2,C}
				\end{bmatrix}
				\begin{bmatrix}
					A \\ 
					B \\
					C
				\end{bmatrix}
				=
				\begin{bmatrix}
					Z_{1,R} \\ 
					Z_{1,L} \\ 
					Z_{1,C}
				\end{bmatrix},
			\end{split}
		\end{equation}
		
		\noindent and performed a matrix inversion to solve for $A$, $B$, and~$C$.  
		
		Step 4, we confirmed that $A$, $B$, and $C$ accurately characterized our system by applying Eq.~\ref{eq:ABCD_subtract} to our measurement of $Z_{1,RLC}$ and overlayed it with its ``truth'' measurement.  
		
		The main limitation of this 1-port calibration method is that the calibration will eventually drift due to changing conditions, e.g. cable temperature.  When this happens, the calibration will need to be repeated.   


	\subsection{Data analysis \label{subsec:analysis}}
		
\begin{figure}
			\includegraphics[]{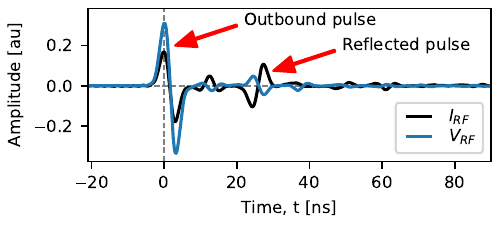}
			\caption{\label{fig:example_RFIV_pulses} A zoomed-in view of a single $I_{RF}(t)$ and $V_{RF}(t)$ pulse.  The outbound monopulses occur at $t=0$ and the reflected pulses are measured 25 ns later.  }
		\end{figure}

		This section details the analysis steps we used to process our raw oscilloscope measurements into plasma properties: density, electron damping, and sheath thickness.  To assist with the discussion, Figure~\ref{fig:example_RFIV_pulses} shows a single $I_{RF}$ and $V_{RF}$ pulse from our dataset in Sec.~\ref{subsec:large_osc}.  
		
		Step 1, we centered each $I_{RF}(t)$ and $V_{RF}(t)$ pulse within its own window of width, $\tau$, and applied a Hann window to suppress content away from the pulses.   
		
		Step 2, we used Eq.~\ref{eq:Z_RFIV} to convert $V_{RF}(t)$ and $I_{RF}(t)$ to impedance, $Z_1(\omega)$, for each window.
		
		Step 3, we applied the 1-port calibration (see Sec.~\ref{subsec:calibrations}) to convert our impedance measurement at plane~1, $Z_1(\omega)$, to plane~2, $Z_2(\omega)$.  Then we applied Eq.~\ref{eq:stem_model_Z3} to  get  $Z_3(\omega)$.
		
		Step 4, we typically fit Eq.~\ref{eq:Z_total2} to our measurement, $Z_3(\omega)$, to get our plasma measurements.  	In our particular dataset (described below), our choice of $\sigma$ was a little low, and this resulted in an intractably low SNR at the higher plasma frequencies.  
		
\begin{figure}
			\includegraphics[]{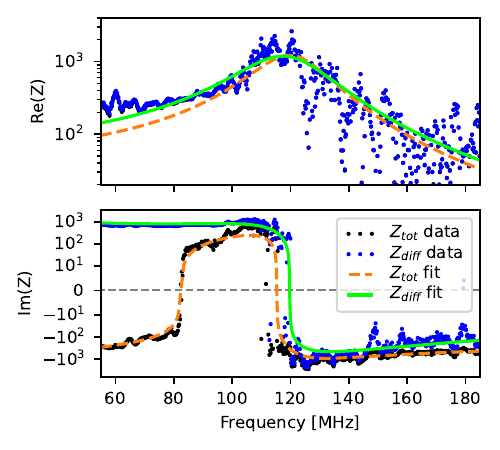}
			\caption{\label{fig:example_fit}  Models of $Z_{tot}$ and $Z_{diff}$ are fit to corrected measurements and show good agreement. Units in Ohms. }
		\end{figure}

		As an alternate step 4, we instead took advantage of the fact that our measurement of $Z_2(\omega)$ contains the same information as $Z_3(\omega)$ but at a lower frequency where SNR is higher.  To this end, we applied Eq.~\ref{eq:stem_model_Z2} to Eq.~\ref{eq:Z_total2} and fit this combined model to our measurement of $Z_2(\omega)$ to get our plasma measurements.  		Figure~\ref{fig:example_fit} shows an example of this fit applied to this dataset which resulted in $f_p = 195$ MHz, $\nu' = 0.185$ ($227 \times 10^6$ s$^{-1}$), $t_{sh}' = 0.149$ (1.08 mm), and $n=4.72 \times 10^{14}$ m$^{-3}$.  A similar procedure was done for $Z_{diff}$ and is also shown.
		
		A second alternative to this 4th step  (not used in this work) is to convert both the measurements and the models to the reflection coefficient~\cite{blackwell_characteristics_2005}
		\begin{equation} \label{eq:gamma}
			\begin{split}
				\Gamma = \frac{Z - Z_0}{Z + Z_0}
			\end{split}
		\end{equation}
		and then fit.  In this equation, $Z_0$ is typically 50 $\Omega$.  This method is recommended and will be discussed in a future publication.


		Finally, steps 1 to 4 were then repeated for each of the sequential monopulses to get time-resolved measurements of $\omega_p$, $\nu$, $t_{sh}$, and $n$.  
		
		
	\subsection{Other considerations}
	
		Poor signal-to-noise ratio (SNR) is one of the biggest limits to time-resolved PIP operation because low SNR increases measurement uncertainty and potentially makes analysis untenable.  Poor SNR is typically caused by poor calibration, calibration drift, electromagnetic interference, etc.  The best method to maximize SNR is to perform careful calibrations and routinely repeat them when drift is observed.  SNR can also be improved by carefully choosing $\sigma$ and $\tau$ as discussed in Sec.~\ref{subsec:setup}.    In postprocessing, SNR can be improved by averaging and using windowing functions (e.g. Hann windows) within each window.  Determining the plasma properties by fitting measurements to data is also more robust to SNR than using the zero intercepts.

\section{Results \label{sec:results}}

	In this section, we apply the time-resolved analysis from Sec.~\ref{sec:methods} to two plasma datasets.   
	The first is characterized by a large-amplitude oscillation and highlights the accessible density range within a single measurement. 
	The second is characterized by a high-frequency (150 kHz) plasma oscillation which we resolve at 4 MHz.  This highlights the improved time resolution achieved in this work.

	\subsection{Large-amplitude dataset \label{subsec:large_osc}}

\begin{figure*}
			\includegraphics[]{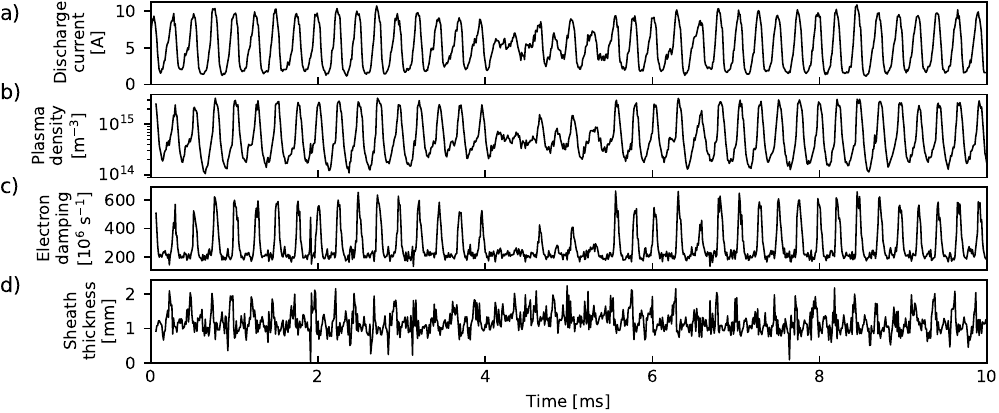}
			\caption{\label{fig:large_amp_osc_results}  Results of the large-oscillation dataset.  a) Time resolved discharge current, b)plasma density, c) electron damping, and d) sheath thickness. }
		\end{figure*}

		The first dataset is characterized by large-amplitude oscillations (50\% of the mean) and relatively low-frequency (4 kHz) in the discharge current.  Because of this frequency, we set the time resolution (i.e. $\tau$) to 10 $\mu$s $=(100 \text{ kHz})^{-1}$.   To improve SNR, each time window in the dataset was processed with a 1~$\mu$s Hann window.  Figure~\ref{fig:large_amp_osc_results} shows the resulting plasma properties.

		Figure~\ref{fig:large_amp_osc_results}a and~b show that the PIP density measurements capture much of the dynamics present in the discharge current, and that density oscillations range from $2\times 10^{14}$ to $3\times 10^{15}$ m$^{-3}$ (an entire order of magnitude).  The density measurement also lags the discharge current by approximately 50~$\mu$s.  Figures~\ref{fig:large_amp_osc_results}c and~d show that the time resolved electron damping and sheath thickness also show similar dynamics as the discharge current, including the relatively quiescent region between 4 and 6 ms.  During this region, the mean sheath thickness increases from roughly 1.1 mm to 1.3 mm.  
	
		This dataset also highlights the importance of choosing $\sigma$ such that the entire range an oscillating $\omega_p(t)$ is measurable (see Figure~\ref{fig:monopulse}b).  In this dataset,  our choice of $\sigma = 1 / (2 \pi 200 \text{ MHz})$ was reasonable for the plasma frequency range (approximately between 100 and 500 MHz), but not optimal because it  also resulted in a relatively low SNR  (i.e. higher measurement uncertainty) at the higher densities.

	\subsection{High-frequency dataset}
	
		The second dataset is characterized by a high-frequency (150 kHz) and relatively low amplitude (7\% of mean) plasma oscillation as measured by the discharge current.  We resolved the PIP measurements at 4~MHz~(1/$\tau$) as a compromise between measurement-amplitude resolution and temporal resolution.  These measurements were processed with a 0.5 $\mu$s Hann window to improve SNR.   Measurements from an AC-coupled ion saturation probe adjacent to the PIP are also provided.
		
		\begin{figure}
			\includegraphics[]{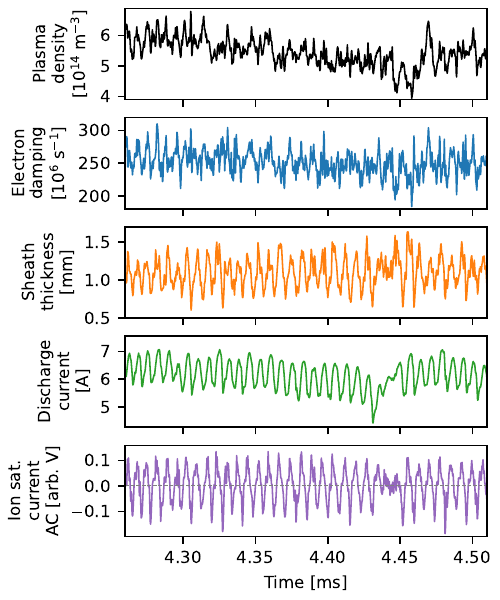}
			\caption{\label{fig:high_freq_osc_results} Results of the high-frequency dataset.  All five measurements show the same 150 kHz oscillation and the same low-frequency ``feature'' at 4.45 ms. The ion saturation units are an uncalibrated voltage. }
		\end{figure}
		
		Figure~\ref{fig:high_freq_osc_results} shows that five time-series measurements show the same high-speed oscillations (150 kHz) and the  lower-frequency ``feature'' at 4.45 ms.  
		
\begin{figure}
			\includegraphics[]{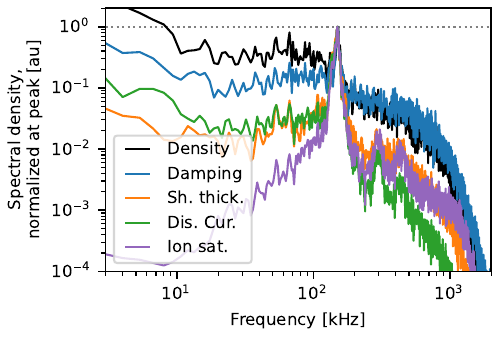}
			\caption{\label{fig:high_freq_osc_fft}  The ratio of the measurement's amplitude at the peak frequency (all at 150 kHz) to the surrounding ``noise floor'' provides an estimate of the SNR for each measurement.  }
		\end{figure}

		Figure~\ref{fig:high_freq_osc_fft} shows the spectral density for each signal normalized by the amplitude at the dominant frequency (150 kHz).  The relative ratio of the peak to surrounding floor provides an approximation of the SNR for each signal, and the results show that density has the lowest SNR followed by the damping rate.    The relatively low SNR is at least partially due to the relatively small amplitude of the plasma oscillations.

\section{Conclusions}
	
	The research discussed above introduced an improved approach to interpreting PIP results and achieving faster time-resolved measurements. These improvements are a result of a simple PIP design (Sec.~\ref{subsec:PIP_design}), improved analytical models (Sec.~\ref{subsec:PIPs_head}), a correction for the PIP's stem (Sec.~\ref{subsec:PIPs_stem}), a numerical error analysis that facilitated a discussion on appropriate probe design (Sec.~\ref{subsec:COMSOL model}), higher-frequency hardware (Sec.~\ref{subsec:setup}), a more rigorous approach to RF calibration (Sec.~\ref{subsec:calibrations}), and updated analysis procedure (Sec.~\ref{subsec:analysis}) including discussing several approaches to improving time-resolved SNR including calibration, parameter optimization, and filtering.  The results (Sec.~\ref{sec:results}) show that density, electron damping, and sheath thickness can be resolved in time for both large oscillations ($2\times 10^{14}$ to $3 \times 10^{15}$~m$^{-3}$) and high-frequency oscillations (150 kHz) and capture many of the same dynamics as accompanying diagnostics.  The maximum time resolution achieved in this work was 4~MHz, but the actual upper limit (related to the electron damping term) was likely between 10 to 50~MHz for our specific plasma conditions.

	Several factors place limits on the effectiveness of this method.  
	First, the PIP models developed in this work make several assumptions, and deviations from these introduce errors to the measurements.  However, these errors can be minimized by making appropriate probe and experimental design choices. 
	Second, time-resolved PIP analysis has a lower SNR (and therefore larger uncertainty) than traditional PIP analysis, primarily due to less time averaging.  We provide several approaches to improving time-resolved SNR through calibration, parameter optimization, and filtering.  


\section*{Acknowledgments}

This analysis was performed while JWB held an NRC Research Associateship award at the Naval Research Laboratory and while MP was a Ph.D. student at the University of Illinois Urbana-Champaign.  Funding was provided by the U.S. Air Force (USAF).

\section*{Data Availability}

The data that support the findings of this study are available from the US Naval Research Laboratory. Restrictions apply to the availability of these data, which were used under license for this study. Data are available from the authors upon reasonable request and with the permission of the US Naval Research Laboratory.

\appendix


\section{Custom PCBs \label{app:pcbs}}

	In this work, we used two custom PCBs and present each below.  

\begin{figure}
		\includegraphics[width=3.33in]{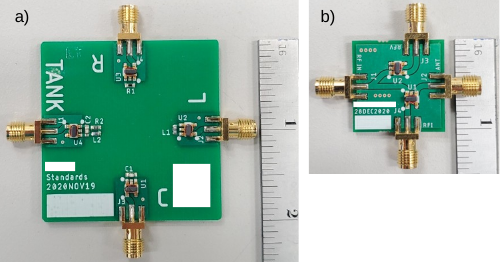}
		\caption{\label{fig:PCBs} a) The ``standards PCB'' contains 3 calibration standards (R, L, C) and an RLC resonant circuit (TANK) used during the 1-port calibration. b) The ``RFIV PCB'' isolates signals proportional to voltage (RFV) and current (RFI) in the PIP's transmission lines.  Units in inches.}
	\end{figure}

	The standards board (Figure~\ref{fig:PCBs}a) is a single PCB that consists of the R, L, and C calibration standards and the RLC resonant tank circuit used to ``sanity check'' the 1-port calibration (Sec.~\ref{subsec:calibrations}).  The RLC resonant circuit is designed to have a resonance around expected plasma frequencies.  The individual circuits are placed far enough apart so that they do not electrically couple, and SMA connectors are used for consistent measurements.  The baluns (transformers) shown in the image are a legacy component and will be removed in future designs. 

	The RFIV board\cite{dubois2021} (Figure~\ref{fig:PCBs}b) is a PCB that uses baluns to isolate signals proportional to the voltage and current that are then measured by the oscilloscope.  
	This PCB should ideally be installed near the PIP (see Figure~\ref{fig:time_series_setup}) to reduce the time between the outgoing and reflected pulses (see Figure~\ref{fig:example_RFIV_pulses} for an example) and therefore prevent sequential pulses from overlapping.

\nocite{*}
\bibliography{biblio}

\end{document}